\documentclass[preprint,showpacs,amsmath,amssymb]{revtex4}
\usepackage{mathrsfs}

\usepackage{graphicx,color}
\usepackage{dcolumn}
\usepackage{bm}
\def\slash#1{{#1\!\!\!/}}
\newcommand{\nb}{\nonumber}
\begin{document}


\title{Next-to-leading order QCD corrections to Higgs boson decay to quarkonium plus a photon}

\author{Zhou Chao$^a$}
\author{Song Mao$^a$}\email{songmao@mail.ustc.edu.cn}
\author{Li Gang$^a$}
\author{Zhou Ya-Jin$^b$}
\author{Guo Jian-You$^a$}

\affiliation{$^a$ School of Physics and Material Science, Anhui University, Hefei, Anhui 230039, P.R.China}
\affiliation{$^b$ School of Physics, Shandong University, Jinan Shandong 250100, P.R. China}

\date{\today}

\begin{abstract}
In this paper, we investigate the decay of Higgs boson to $J/\psi(\Upsilon)$ plus a photon based on NRQCD factorization. For the direct process, we calculate the decay width up to  QCD NLO.  We find that the decay width for process $H \to J/\psi(\Upsilon)+ \gamma$ direct production at the LO is significantly reduced by the NLO QCD corrections. For the indirect process, we calculate the $H \to \gamma^\ast\gamma$ with virtual $\gamma$ substantially decaying to $J/\psi(\Upsilon)$, including all the SM Feynman diagrams.
The decay width of indirect production is much larger than the direct decay width. Since it is very clean in experiment, the $H \to J/\psi(\Upsilon)+ \gamma$ decay could be observable at a 14 TeV LHC and it also offers a new way to probe the Yukawa coupling and New Physics at the LHC.
\end{abstract}

 \maketitle

\section{Introduction}
\par
Recently, both ATLAS and CMS collaborations have announced that they
observed a new boson with mass around 125 GeV, whose properties are
consistent with the Standard Model(SM) Higgs in any measured channel
\cite{higgs1,higgs2,higgs3,higgs4}. After discovery of the Higgs
boson, the main task is to determine its properties, such as spin,
CP, and couplings. The couplings to gauge bosons and the third-generation fermions are measured
directly, which are fixed through the well measured diboson decays of
the Higgs determined at the $20\sim30\%$ level. However, we have little information about the Higgs Yukawa
couplings to the first- and second-generation quarks at current experiments,
since these couplings are predicted to be small in the SM,
and the inclusive decays of the Higgs to these states are swamped by large QCD backgrounds.
These couplings are indirectly and weakly constrained by the inclusive Higgs production cross
section \cite{coupling1,coupling2}. Such constraints only probe the simultaneous deviation of all Yukawa couplings.
They do not provide information about the separate Yukawa couplings of the different quarks.

\par
The study of heavy quarkonium is one of the interesting
subjects in high energy physics, which offers
a good testing ground for investigating the Quantum Chromodynamics (QCD)
in both the perturbative and non-perturbative regimes.
The factorization formalism of non-relativistic QCD (NRQCD) \cite{bbl} as a
rigorous theoretical framework to describe the heavy-quarkonium
production and decay has been widely investigated both at experimental and theoretical aspects.
Many experimental data of the heavy quarkonium production and decay are fairly well described by the NRQCD theory\cite{d1,d2,comtev,chao2011,pnloz,Butenschoen:1105}.

\par
Recent works showed that the exclusive decays of the Higgs boson to vector
mesons can probe the Yukawa couplings of first- and second-generation quarks at future runs of the LHC \cite{Bodwin:2013gca}.
These couplings are hard to access in hadron colliders
through the direct $H \to q\bar{q}$ decays, owing to the overwhelming QCD background.
While the Yukawa couplings $Hc\bar{c}$ might be probed at the LHC by making use of
charm-tagging techniques, its phase must be determined through the processes involving
quantum interference effects, such as the decay \cite{Delaunay:2013pja}. Although the branch ratios of Higgs boson to vector
mesons are small, it offers complimentary information about Higgs couplings and can serve
as searching New Physics (NP) beyond SM. Besides, subsequent decays of $J/\psi(\Upsilon)$ into pair of leptons is a clean channel in experiments.
Recently, Higgs rare decay to a vector quarkonium ($J/\psi,\Upsilon$) received considerable attention \cite{Kagan:2014ila,Gao:2014xlv,Modak:2014ywa,Delaunay:2013pja,Colangelo:2016jpi}.
The relativistic correction for Higgs boson decay to an S-wave vector
quarkonium plus a photon have been calculated in Ref. \cite{Bodwin:2014bpa}.
A search for the decays of the Higgs and Z bosons to $J/\psi$ and $\Upsilon$ is performed
in integrated luminosities $20.3 fb^{-1}$ with the ATLAS detector at 8 TeV LHC. No significant
excess of events is observed above expected backgrounds and $95\%$ CL upper limits are placed on the
branching fractions. In the $J/\psi \gamma$ and $\Upsilon(1S)\gamma$
final state the limits are $1.5\times10^{-3}$ and $1.3\times10^{-3}$  for the Higgs boson, respectively \cite{Aad:2015sda}.

\par
As we know, the NLO QCD corrections to quarkonium production are
usually significant \cite{nloqcd,nlo1,nlo2}. We should generally
take the NLO QCD corrections into account in studying heavy-quarkonium
production processes. In this paper, we will calculate the
$H \to J/\psi(\Upsilon) + \gamma$ process up to the QCD NLO
within the NRQCD framework by applying the covariant
projection method \cite{p2}. The paper is organized as follows:
we present the details of the calculation strategies in Sec.II.
The numerical results are given in Sec.III. Finally, a short summary and discussions
are given.

\section{Calculation descriptions}
\subsection{LO calculation for direct production }
\par
\begin{figure}[!htb]
\begin{center}
\begin{tabular}{cc}
{\includegraphics[width=0.6\textwidth]{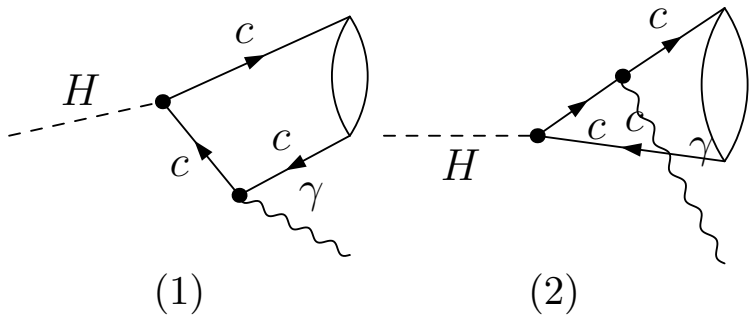}}
\end{tabular}
\end{center}
 \vspace*{-0.7cm}\caption{The Feynman diagrams for the $H\to c\bar{c}[^3S_1^{[1]}]+ \gamma $ direct decay process at the LO} \label{f1}
\end{figure}

We begin to discuss the decay $H \to J/\psi + \gamma$. Since the
calculation of the $\Upsilon$ decay is identical to the $J/\psi$, we
will not present it explicitly in this section. There are two
Feynman diagrams for this process at leading order(LO), which are
shown in Fig.\ref{f1}. We calculate the amplitudes by making
use of the standard methods of NRQCD factorization \cite{bbl}. The
process $H\to c\bar{c}+ \gamma $ at LO is denoted as:
\begin{equation}\label{emission}
H(p_1)\to c(p_2)\bar{c}(p_3)+\gamma(p_4).
\end{equation}
The amplitudes for the two diagrams are given by
\begin{eqnarray}
{\cal M}_{i1} =  \bar{u}(p_2)\cdot\frac{-i e m_c}{2m_W s_W}\cdot \frac{i}{\rlap /p_1-\rlap /p_2-m_c}\cdot i\frac{2}{3}e \gamma^\mu \cdot v(p_3)
                \epsilon_{\mu}^*(p_4),~ \nonumber \\
{\cal M}_{i2} = \bar{u}(p_2) \cdot i\frac{2}{3}e \gamma^\mu \cdot \frac{i}{\rlap /p_1-\rlap /p_3-m_c} \cdot\frac{-i e m_c}{2m_W s_W}\cdot v(p_3)
                \epsilon_{\mu}^*(p_4).~ \nonumber \\
\end{eqnarray}
The relative momentum between the $c$ and $\bar{c}$ is defined as $q
= (p_2 - p_3)/2$, and the total momentum of the $J/\psi$ is defined
as $p = p_2 + p_3$. Then, we obtain the following relations among
the momenta:

\begin{eqnarray}
p_{2} &=& \frac{1}{2}p + q, \;\;\; p_{3}= \frac{1}{2}p - q,  \;\;\; p \cdot q = 0,\nonumber \\
p^{2}_{2} &=& p^{2}_{3}=m^{2}_{c}, \;\;\; p^2 = E^2,\;\;\; q^2 =
m_c^2-E^2=-m_c^2 v^2. \label{eq:momdefs}
\end{eqnarray}
In the $c\bar{c}$ rest frame, $p = (E, 0)$ and $q = (0, q)$. In the non-relativistic $v=0$ limit, $p^2 = 4 m^2_c,~~ q^2 = 0$. In
order to produce a $J/\psi$, the $c\bar{c}$ pair must be produced in
a spin-triplet, color-singlet Fock state. We can obtain the
short-distance amplitudes by applying certain projectors onto the
usual QCD amplitudes for open $c\bar{c}$ production. By using the
notations in Ref.\cite{p2}, we get the amplitudes:

\begin{eqnarray}
{\cal M}_{ ^3S_1^{(1)} } = {\cal E}_{\alpha} {\rm Tr}
\Big[ {\cal C}_{1} \Pi_{1}^{\alpha} {\cal M} \Big]_{q=0}, \nonumber
\end{eqnarray}
where the spin-triplet projector is given by
\begin{eqnarray}
&&{\mit \Pi}_{1}^{\alpha}
     = {1\over{\sqrt{8m^3}}} \left({{\slash{p}}\over 2} - \slash{q} - m\right)
       \gamma^\alpha \left({{\slash{p}}\over 2} + \slash{q} + m\right)\, .
\end{eqnarray}

The colour singlet state will be projected out by contracting the amplitudes with the following
operators :
\begin{eqnarray}
&&{\cal C}_1 = {{\delta_{ij}}\over{\sqrt{N_c}}}
\label{proj_sing}
\end{eqnarray}
The amplitude ${\cal M}$ is obtained by calculating the two Feynman diagrams in Fig.\ref{f1} in QCD
perturbation theory. The trace is over both the Lorenz and color indices.

After the application of this set of rules, we obtain the short-distance partial decay width
$\hat \Gamma$ for $H\to c\bar{c}[^3S_1^{[1]}]+ \gamma $  processes:
\begin{eqnarray}
&&d\hat\Gamma(H \to c\bar{c}[^3S_1^{[1]}]+ \gamma) = {1\over{32 \pi^2}}
     |{\cal M}_{^3S_1^{[1]}}|^2\;{|p|\over{m_H^2}}d \Omega \, ,
\end{eqnarray}
where $|p|=\frac{m_H^2-m^2_{J/\psi}}{2 m_H}$  and $m_H$ represent the Higgs boson
mass. $d\Omega=d\phi d(cos\theta)$ is the solid angle of particle $J/\psi$.
\begin{eqnarray}
&&{|{\cal M}_{^3S_1^{[1]}}|^2 = {{256 \pi^2 \alpha^2 m_c}\over{3 m_W^2s_W^2}}}
\label{proj_sing}
\end{eqnarray}
The decay width read:

\begin{eqnarray}
\Gamma (H\to J/\psi+\gamma) =  \hat{\Gamma}(H\to c\bar{c}(^3S_1^{[1]})+\gamma) \frac{< {\cal{O}}^{J/\psi}(^3S_1^{[1]})>}{2 N_c N_{col} N_{pol}},
\end{eqnarray}
where $N_{col}$ and $N_{pol}$ refer to the number of colours and polarization states
of the  $c\bar{c}$ pair produced.  The color-singlet
states $N_{col}=1$, and $N_J=3$ for polarization vectors $^3S_1^{[1]}$ state in 4 dimensions.
$2 N_c$ is due to the difference between the conventions in Ref.~\cite{p2} and Ref.~\cite{bbl}.

\subsection{NLO Calculation for direct production}

\par
\begin{figure}[!htb]
\begin{center}
\begin{tabular}{cc}
{\includegraphics[width=0.6\textwidth]{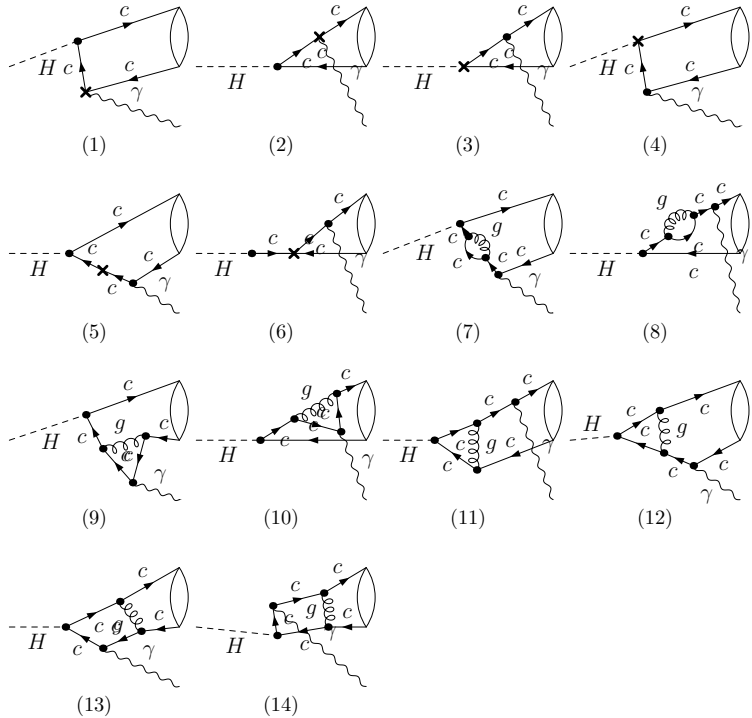}}
\end{tabular}
\end{center}
 \vspace*{-0.7cm}\caption{The Feynman diagrams for the $H\to c\bar{c}[^3S_1^{[1]}]+ \gamma $ decay process at QCD NLO.} \label{f2}
\end{figure}

At LO, we have two contributing Feynman diagrams, and at NLO there are 14.  All Feynman
diagrams are generated with program FeynArts, and evaluated by using our in-house program, which is written in the programming language Mathematics.
The one-loop diagrams for the ${\cal{O}}(\alpha_s)$ corrections to $H\to c\bar{c}[^3S_1^{[1]}]+ \gamma $
are shown in Fig.\ref{f2}. The $\alpha_s$ corrections involving virtual gluons include the interferences between Born diagrams and one-loop virtual diagrams.
The virtual diagrams are computed analytically
and all tensor integrals are reduced to linear combinations of one-loop scalar functions.
The virtual corrections contain Ultraviolet (UV), Infrared (IR) and Coulomb singularities.
In our calculations, we adopt the dimensional regularization
(DR) scheme to regularize the UV and IR
divergences in $D$ dimensions with $D \equiv 4 - 2\epsilon$. The UV singularities of the virtual corrections are removed by introducing
a set of related counterterms.
The counterterms for the charm quark wave function and the charm quark mass are defined as
\begin{eqnarray}
\psi^{0}_{c} & = & \left(1+\frac{1}{2}\delta Z_{c}\right)\psi_{c}~, \\  \nonumber
m^{0}_{c} & = & m_{c} +  \delta m_{c} ~,
\end{eqnarray}
The on-mass-shell scheme is adopted
to fix the wave function and mass renormalization constant of the external
charm quark field, then we obtain
\begin{eqnarray}
\delta Z_{c} & = & - 3 C_F \frac{\alpha_s}{4\pi} \left[\Delta_{UV} + \ln \frac{\mu_r^2}{m_c^2} +\frac{4}{3} \right]~, \\ \nonumber
\frac{\delta m_{c}}{m_{c}} & = & - \frac{\alpha_s }{3\pi} \left[3 \Delta_{UV} + 4 + \ln \frac{\mu_r^2}{m_c^2} \right]~,
\end{eqnarray}
where $\Delta_{UV}=\frac{1}{\epsilon_{UV}}-\gamma_E + \ln (4\pi)$.
After applying the renormalization procedure the UV
divergences in the virtual correction are canceled.
The IR singularities are analytically canceled when we
added all the virtual Feynman diagrams together.
We adopt the expressions in Ref.\cite{IRDV} to deal with the IR
divergences in Feynman integral functions, and apply the expressions
in Refs.\cite{OneTwoThree,Four,Five} to implement the numerical
evaluations for the IR safe parts of N-point integrals. In the
virtual correction calculation, we find that only Fig.\ref{f2}(13) and
Fig.\ref{f2}(14) induce Coulomb singularities, and we use a small relative
velocity $v$ between $c$ and $\bar{c}$ to regularize them
\cite{coulomb}.

\subsection{Indirect Decay Calculation}
\begin{figure}[!htb]
\begin{center}
\begin{tabular}{cc}
{\includegraphics[width=0.8\textwidth]{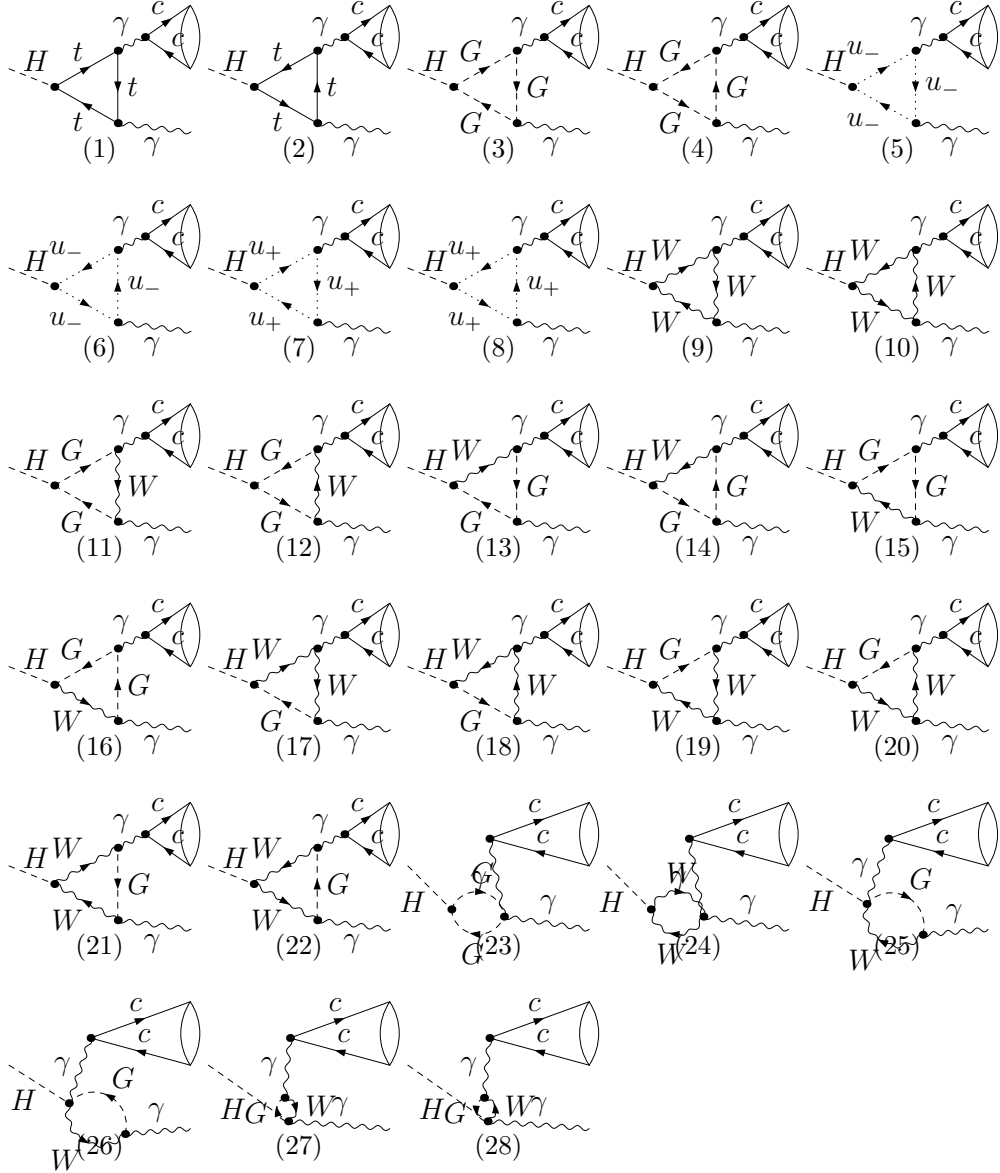}}
\end{tabular}
\end{center}
 \vspace*{-0.7cm}\caption{The Feynman diagrams for the $H\to c\bar{c}[^3S_1^{[1]}]+ \gamma $ indirect decay process.} \label{f3}
\end{figure}

The direct Higgs decay process to the heavy quarkonium plus photon, is mainly produced through the Higgs and charm quarks Yukawa coupling.
While the indirect decay process is mainly produced through Higgs decaying into two photons, then one virtual photon substantially decaying to a $c\bar{c}$ quark pair. Since Higgs decays into di-photon process is forbidden at tree level in SM, the leading order contribution comes from the one-loop Feynman diagrams,
including top quark and $W$ boson triangle diagrams, which are shown in Fig.\ref{f3}. Due to the fact that the coupling strength of Higgs and top($W$) is proportional to the particle mass, the contribution of indirect decay is not small. The process Higgs decays into di-photon at leading order in $\alpha_s$ have been calculated in Ref.\cite{two-photon}. The two-loop electroweak and QCD corrections to this process have also been studied in Ref. \cite{two-loop}. In Ref. \cite{Bodwin:2013gca}, the authors gave the approximate results for the Higgs decay to $J/\psi(\Upsilon)$ and photon through Higgs decaying into two photon.  In our paper, we analytically calculate this process based on NRQCD factorization. In Feynman gauge, there are 28 Feynman diagrams, which include the contribution from not only the top, W-boson loops, but also the ghost and goldstone loops.
First we generate the amplitudes of Higgs decay to di-photon, which is given by
\begin{eqnarray}
{\cal M}_{H \to \gamma\gamma}^{\mu\nu} = \frac{i \alpha^\frac{3}{2}}{24 m_W s_W \sqrt{\pi}}\times({\cal A} g^{\mu\nu}+{\cal B} p^{\nu}p_4^{\mu})~ \nonumber \\
\end{eqnarray}

The expressions of coefficients ${\cal A}$ and ${\cal B}$ are listed in the appendix. Then we multiply it to the amplitude of virtual photon decay to $c\bar{c}$ quarks pair.
After the application of the projection operator, we get the short-distance amplitude,
\begin{eqnarray}
{\cal M}_{indirect} =  {\cal M}_{H \to \gamma^\ast\gamma}^{\mu\nu} \frac{2}{3}e \frac{- g^{\mu\sigma}}{p_1^2} {\rm Tr} \Big[ {1\over{\sqrt{8m_c^3}}} \left({{\slash{p}}\over 2} - \slash{q} - m_c\right)
       \gamma^\mu \left({{\slash{p}}\over 2} + \slash{q} + m_c\right) {{\delta_{ij}}\over{\sqrt{N_c}}} \Big]_{q=0} \epsilon^\ast_{\sigma}\epsilon^\ast_{\nu},~
\end{eqnarray}
Following the Passarino-Veltman(PV) method [14, 16], we can expressed the tensor integrals
as a linear combination of tensor structures and coefficients, where the tensor
structures depend on the external momenta and the metric tensors, while the coefficients
depend on one-loop scalar integrals, kinematics invariants and the dimension of the integral.
The one-loop integrals are calculated analytically by using dimensional regularization in $D=4-2\epsilon$ dimensions.
Finally, we squared on all the amplitudes in 4 dimensions.

\section{Numerical results and discussion}
\par
In this section, we discuss our numerical results for both the central values and theoretical
errors for the $H\to J/\psi+\gamma$ and $H\to \Upsilon(1S)+\gamma$ decays.  The relevant parameters are taken as follows \cite{pdg}:
\begin{eqnarray}
&\alpha^{-1}&=137.036,~m_Z = 91.1876~{\rm GeV},~ m_W = 80.385~{\rm GeV},~m_c=m_{J/\psi}/2=1.5~{\rm GeV}, \nonumber \\
&m_b&=m_{\Upsilon(1S)}/2=4.75~{\rm GeV}, ~m_t=173.2~{\rm GeV}.
\end{eqnarray}
We take two-loop running $\alpha_s$ in the calculation, and the
corresponding fitted value $\alpha_s(M_Z) = 0.118$ is used for the calculations. The
renormalization and NRQCD scales are chosen as
$\mu_r=m_H$ and $\mu_{\Lambda}=m_c(m_b)$, respectively.
The Long Distance Matrix Elements (LDME) for $J/\psi$ and $\Upsilon$ used in this paper are set as:
\cite{kniehl1201}
 $<{\cal O}^{J/\psi}[^3S_1^{(1)}]>= 1.3 ~{\rm GeV^3} $,
 $<{\cal O}^{\Upsilon}[^3S_1^{(1)}]>= 9.28 ~{\rm GeV^3} $.

Finally, we get the decay widths for $H\to J/\psi(\Upsilon)+\gamma$ for the direct and indirect processes:

\begin{eqnarray}
\Gamma_{LO}^{direct}(H\to J/\psi+\gamma) =5.334\times10^{-10}~ {\rm GeV}\nb \\
\Delta\Gamma_{NLO}^{direct}(H\to J/\psi+\gamma) =-4.099\times10^{-10}~ {\rm GeV} \nb \\
\Gamma^{direct}(H\to J/\psi+\gamma) =1.235\times10^{-10}~{\rm GeV} \nb \\
\Gamma^{indirect}(H\to J/\psi+\gamma) =1.013\times10^{-7}~ {\rm GeV} \nb \\
\Gamma_{LO}^{direct}(H\to \Upsilon+\gamma) =2.998\times10^{-9}~ {\rm GeV} \nb \\
\Delta\Gamma_{NLO}^{direct}(H\to \Upsilon+\gamma) =-1.845\times10^{-9}~ {\rm GeV} \\
\Gamma^{direct}(H\to \Upsilon+\gamma) =1.153\times10^{-9}~ {\rm GeV} \nb \\
\Gamma^{indirect}(H\to \Upsilon+\gamma) =5.659\times10^{-9}~ {\rm GeV}
\end{eqnarray}

Using the width of Higgs boson decays within Standard Model $\Gamma(H) = 4.195^{+0.164}_
{-0.159} \times {10^{-3}}~ {\rm GeV}$\cite{higgs-decay}, we obtain the following results for the branching
fractions in the SM:

\begin{eqnarray}
{\cal B}^{direct}(H\to J/\psi+\gamma) =2.94\times10^{-8}~ \nb \\
{\cal B}^{indirect}(H\to J/\psi+\gamma) =2.41\times10^{-5}~\nb \\
{\cal B}^{direct}(H\to \Upsilon+\gamma) =2.75\times10^{-7}~ \nb \\
{\cal B}^{indirect}(H\to \Upsilon +\gamma) =1.35\times10^{-6}~
\end{eqnarray}.

As the results show, for the decay process $H\to J/\psi+\gamma$, the direct contribution is much smaller than the indirect contribution,
so it is difficult to observe direct contribution in the total cross section and not suitable for studying the coupling of Higgs and charm quarks.
For the process $H\to \Upsilon+\gamma$, the direct and indirect contributions are comparable and the total cross section is sensitive to the direct decay process, so this process can be used to study the coupling of Higgs and bottom quarks.
In addition, the process $H\to J/\psi(\Upsilon)+\gamma$ decay can also be used to test the coupling of Higgs to top quarks and W boson.

The main uncertainties for the results of $H\to J/\psi(\Upsilon)+\gamma$ arise from the uncertainties in LDMEs, renormalization scale,
and the relativistic corrections. The relativistic corrections and the uncertainties have been discussed in Ref. \cite{Bodwin:2014bpa}.
In Table.\ref{tab:mu}, we illustrate the renormalization scale dependence of the direct and indirect decay widths for the process $H\to J/\psi(\Upsilon)+\gamma$. We assume $\mu=\mu_r$ and define $\mu_0 = m_H $. When the scale $\mu$ running from $\mu_0/4$ to $4 \mu_0$,
The related theoretical uncertainty for $H\to J/\psi+\gamma$ amounts to $^{+55.0}_{-83.2}\%$ for direct process and to $^{+4.2}_{-4.1}\%$ for indirect process, and the related theoretical uncertainty for $H\to \Upsilon+\gamma$ amounts to $^{+40.2}_{-26.5}\%$ for direct process and to $^{+4.2}_{-4.2}\%$ for indirect process. The LO direct process is independent of the renormalization scale $\mu_R$, because it is pure electroweak channels.

\begin{table}[t]
  \centering
  \begin{tabular}{c|c|c|c|c|c|c|c|c}
  \hline\hline
 & \multicolumn{4}{|c|}{$H\to J/\psi+\gamma ~~(\times10^{-10}~{\rm GeV})$}& \multicolumn{4}{|c}{$H\to \Upsilon+\gamma ~~(\times10^{-10}~{\rm GeV})$ } \\
\hline
$\mu$ [GeV] &  $\Gamma_{LO}^{direct}$ & $\Delta\Gamma_{NLO}^{direct}$& $\Gamma^{direct}$ & $\Gamma^{indirect}$ & $\Gamma_{LO}^{direct}$ & $\Delta\Gamma_{NLO}^{direct}$& $\Gamma^{direct}$ & $\Gamma^{indirect}$  \\
\hline
 $\mu_0/4$ & 5.334 & -5.127 & 0.207 & 1056  & 29.98 & -23.08  & 6.9  & 58.98  \\
 $\mu_0/2$ & 5.334 & -4.554 & 0.780 & 1035 & 29.98  & -20.50  & 9.48 & 57.78 \\
 $\mu_0$ & 5.334 & -4.099 & 1.235 & 1013 & 29.98  & -18.45  & 11.53  & 56.59\\
 $2 \mu_0$ & 5.334 & -3.728 & 1.606 & 992.1 & 29.98  & -16.79 & 13.19  & 55.41 \\
 $4 \mu_0$ & 5.334 & -3.420 & 1.914 & 971.3 & 29.98  & -15.40  & 14.58  & 54.24 \\
\hline\hline
\end{tabular}
\caption{The renormalization scale dependence of the direct and indirect decay widths for the process $H\to J/\psi(\Upsilon) +\gamma$} \label{tab:mu}
\end{table}

\section{Summary}
In this paper, we investigated the decay of Higgs boson to $J/\psi(\Upsilon)$
plus a photon based on NRQCD factorization. For the direct process, we have calculated the decays width up to QCD NLO and found that
the LO decay widths are significantly reduced by the NLO QCD corrections.
For the indirect process, we calculated the process $H \to \gamma^\ast\gamma$ with virtual $\gamma$ substantially decaying to $J/\psi(\Upsilon)$, including all the SM diagrams. The decay width of indirect production is much larger than the direct decay width. Therefore, it is difficult to probe the Yukawa coupling of Higgs and charm quarks using the process $H\to J/\psi(\Upsilon) +\gamma$. However, it still offers a new way to probe the Yukawa coupling of Higgs and top quarks or bottom quarks and New Physics at the LHC.

\section{Acknowledgments}
This work was supported in part by the National Natural Science Foundation of China (No.11305001,
No.11105083) and financed by the 211 Project of Anhui University (No.02303319).

\begin{appendix}
\setcounter{equation}{0}
\section*{Appendix}

In this Appendix, we list the expression of coefficients ${\cal A}$ and ${\cal B}$ for process Higgs indirect decay to di-photon, respectively. The one-loop integrals are defined as in Ref.\cite{looptools}.
Our results are shown as follows,

\begin{eqnarray*}
{\cal A} &=& -32 m_T^2 + 81 m_W^2 + 12 m_W^2 \ast B0[0, m_W^2, m_W^2]  \\
  & + & 12 m_W^2 B0[4m_C^2, m_W^2, m_W^2] - 64m_T^2 \ast B0[m_H^2, m_T^2, m_T^2]  \\
  & + &   12 m_H^2 B0[m_H^2, m_W^2, m_W^2] + 72 m_W^2 \ast B_0[m_H^2, m_W^2, m_W^2]   \\
  & +&   128 m_C^2 m_T^2 \ast C0i[cc0, 4 m_C^2, m_H^2, 0, m_T^2, m_T^2, m_T^2]   \\
  & -&   32 m_H^2 m_T^2 \ast C0i[cc0, 4 m_C^2, m_H^2, 0, m_T^2, m_T^2, m_T^2]  \\
  & -&   216 m_C^2 m_W^2 \ast C0i[cc0, 4 m_C^2, m_H^2, 0, m_W^2, m_W^2, m_W^2]   \\
  & +&   54 m_H^2 m_W^2 \ast C0i[cc0, 4 m_C^2, m_H^2, 0, m_W^2, m_W^2, m_W^2]   \\
  & +&   24 m_W^4 \ast C0i[cc0, 4 m_C^2, m_H^2, 0, m_W^2, m_W^2, m_W^2]   \\
  & +&   256 m_T^2 \ast C0i[cc00, 4 m_C^2, m_H^2, 0, m_T^2, m_T^2, m_T^2]   \\
  & -&   48 m_H^2 \ast C0i[cc00, 4 m_C^2, m_H^2, 0, m_W^2, m_W^2, m_W^2]   \\
  & -&  378 m_W^2 \ast C0i[cc00, 4 m_C^2, m_H^2, 0, m_W^2, m_W^2, m_W^2]   \\
  & -&   120 m_C^2 m_W^2 \ast C0i[cc1, 4 m_C^2, m_H^2, 0, m_W^2, m_W^2, m_W^2]   \\
  & +&    18 m_H^2 m_W^2 \ast C0i[cc1, 4 m_C^2, m_H^2, 0, m_W^2, m_W^2, m_W^2]   \\
  & -&  264 m_C^2 m_W^2 \ast C0i[cc11, 4 m_C^2, m_H^2, 0, m_W^2, m_W^2, m_W^2]   \\
  & - &    264 m_C^2 m_W^2 \ast C0i[cc12, 4 m_C^2, m_H^2, 0, m_W^2, m_W^2, m_W^2]   \\
  & + &   66 m_H^2 m_W^2 \ast C0i[cc12, 4 m_C^2, m_H^2, 0, m_W^2, m_W^2, m_W^2]   \\
  & - &    24 m_C^2 m_W^2 \ast C0i[cc2, 4 m_C^2, m_H^2, 0, m_W^2, m_W^2, m_W^2]  \\
  & + &   6 m_H^2 m_W^2 \ast C0i[cc2, 4 m_C^2, m_H^2, 0, m_W^2, m_W^2, m_W^2],
\end{eqnarray*}

\begin{eqnarray*}
{\cal B} &=& 2 \times (32 m_T^2 \ast C0i[cc0, 4 m_C^2, m_H^2, 0, m_T^2, m_T^2, m_T^2]  \\
    &-& 12 m_W^2 \ast C0i[cc0, 4 m_C^2, m_H^2, 0, m_W^2, m_W^2, m_W^2]   \\
    &-&  128 m_T^2 \ast C0i[cc12, 4 m_C^2, m_H^2, 0, m_T^2, m_T^2, m_T^2]   \\
    &+& 24 m_H^2 \ast C0i[cc12, 4 m_C^2, m_H^2, 0, m_W^2, m_W^2, m_W^2]   \\
    &+& 57 m_W^2 \ast C0i[cc12, 4 m_C^2, m_H^2, 0, m_W^2, m_W^2, m_W^2]   \\
    &+&  18 m_W^2 \ast C0i[cc2, 4 m_C^2, m_H^2, 0, m_W^2, m_W^2, m_W^2]).
\end{eqnarray*}

\end{appendix}


\end{document}